\documentclass[12pt]{article} 
\usepackage{graphicx,subfigure}
\usepackage{amsmath}
\usepackage{amsfonts}
\newcommand{\be}{\begin{equation}}
\newcommand{\ee}{\end{equation}}
\newcommand{\bear}{\begin{eqnarray}}
\newcommand{\eear}{\end{eqnarray}}
\newcommand{\ba}{\begin{array}}
\newcommand{\ea}{\end{array}}

\begin{document}

\begin{titlepage}
\vfill
\begin{flushright}
{\normalsize BNL-94527-2010-JA}
\end{flushright}

\vfill
\begin{center}
{\Large\bf  Chiral Magnetic Wave}

\vskip 0.3in

Dmitri E. Kharzeev$^{1,2}$\footnote{e-mail:
{\tt dmitri.kharzeev@stonybrook.edu}}
 and 
Ho-Ung Yee$^{1}$\footnote{e-mail:
{\tt hyee@tonic.physics.sunysb.edu}}
\vskip 0.15in

 {\it $^{1}$Department of Physics and Astronomy, Stony Brook University,} \\
{\it Stony Brook, New York 11794-3800 }\\[0.3in]
{\it $^{2}$Department of Physics, Brookhaven National Laboratory,} \\
{\it Upton, New York 11973-5000 }\\[0.3in]

{\normalsize  December 29, 2010}

\end{center}

\vfill

\begin{abstract}
We consider a relativistic plasma containing charged chiral fermions in an external magnetic field, e.g a chirally symmetric quark-gluon plasma created in relativistic heavy ion collisions. We show that triangle anomalies imply the existence of a new type of collective gapless excitation in this system that stems from the coupling between the density waves of the electric and chiral charges; we call it "the Chiral Magnetic Wave" (CMW). The CMW exists even in a neutral plasma, i.e. in the absence of the axial and vector chemical potentials. We demonstrate the existence of CMW and study its properties using three different approaches: i) relativistic magnetohydrodynamics; ii) dimensional reduction to $(1+1)$ Sine-Gordon model, appropriate in a strong magnetic field; and iii) holographic QCD (Sakai-Sugimoto model), appropriate at strong coupling. We also briefly discuss the phenomenological implications of the CMW for heavy ion collisions.
\end{abstract}

\vfill

\end{titlepage}
\setcounter{footnote}{0}

\baselineskip 18pt \pagebreak
\renewcommand{\thepage}{\arabic{page}}
\pagebreak

\section{Introduction}

Recently, the r\^ole of triangle anomalies in the dynamics of relativistic plasmas in magnetic field and/or at finite angular momentum has excited considerable attention. Such plasmas are created for example in relativistic heavy ion collisions at RHIC and LHC where the initial energy density significantly exceeds the threshold for the production of decofined and chirally symmetric quark-gluon plasma, and the coherent electromagnetic fields of colliding ions create a pulse of very intense magnetic field. Of particular interest are the following two phenomena caused in the quark-gluon plasma by the axial anomaly: 
the Chiral Magnetic Effect (CME) and the Chiral Separation Effect (CSE). 

The CME is the phenomenon of electric charge separation along the axis of the applied magnetic field in the presence of fluctuating topological charge \cite{Kharzeev:2004ey,Kharzeev:2007tn,Kharzeev:2007jp,Fukushima:2008xe,Kharzeev:2009fn}. The CME in QCD coupled to electromagnetism assumes a chirality asymmetry
between left- and right-handed quarks, parametrized by an axial
chemical potential $\mu_A$.  Such an asymmetry can arise if there is
an asymmetry between the 
topology-changing transitions early in the heavy ion
collision.  
In particular, at finite axial chemical potential $\mu_A$, an external magnetic field induces the vector current $j_i = \bar{\psi} \gamma_i \psi$:
\be
\vec j_V={N_c\ e \over 2\pi^2} \mu_A \vec B; \label{cme}
\ee
in our present convention the current of electric charge is $e j_V$.
Closely related phenomena have been discussed in the physics of
primordial electroweak plasma \cite{Giovannini:1997gp} and quantum
wires \cite{acf}.  While the original derivation used the weak coupling methods, the origin of the effect is essentially topological and so the CME is not renormalized even at strong coupling, as was shown by the holographic methods  \cite{Yee:2009vw,Rubakov:2010qi,Rebhan:2009vc,Gynther:2010ed,Gorsky:2010xu,Brits:2010pw}. The evidence for the CME has been found in lattice QCD coupled to electromagnetism, both within the quenched approximation \cite{Buividovich:2009wi,Buividovich:2009zzb,Buividovich:2010tn} and with light domain wall fermions \cite{Abramczyk:2009gb}. 

Recently, STAR \cite{:2009uh, :2009txa} and PHENIX \cite{phenix,Ajitanand:2010rc}
Collaborations at Relativistic Heavy Ion Collider reported
experimental observation of charge asymmetry fluctuations.  While the
interpretation of the observed effect is still under intense
discussion, the fluctuations in charge asymmetry have been predicted \cite{Kharzeev:2004ey}
to occur in heavy ion collisions due to the CME. Additional tests include the correlation between the electric and baryon charge asymmetries \cite{Kharzeev:2010gr}. There is an active ongoing discussion of the microscopic mechanisms of CME \cite{Nam:2009jb,Fukushima:2010vw,Orlovsky:2010ga,Zhitnitsky:2010zx,Gorsky:2010dr,KerenZur:2010zw,Fu:2010rs} and of the quantitative estimates of the expected charge asymmetries and of possible backgrounds -- see e.g. \cite{Skokov:2009qp,Bzdak:2009fc,Fukushima:2009ft,Voloshin:2010ut, Muller:2010jd,Mages:2010bc,Rogachevsky:2010ys,Schlichting:2010na,Toneev:2010ph}. 

The Chiral Separation Effect (CSE) refers to the separation of chiral charge along the axis of external magnetic field at finite density of vector charge (e.g. at finite baryon number density) \cite{son:2004tq,Metlitski:2005pr,son:2009tf}. The resulting axial current is given by 
\be
\vec j_A={N_c\ e\over 2\pi^2} \mu_V \vec B , \label{cse}
\ee
where $\mu_V$ is the vector chemical potential. 
The close connection between CME and CSE can be established for example by the method of dimensional
reduction appropriate in the case of a strong magnetic field
\cite{Basar:2010zd}: the simple relations $J_V^0 = J_A^1, \ J_A^0 =
J_V^1$ between the vector $J_V$ and axial $J_A$ currents in the
dimensionally reduced $(1+1)$ theory imply that the density of
baryon charge must induce the axial current, and the density of
axial charge must induce the current of electric charge (CME); see also Ref.\cite{holog_spiral}.
Since in the strong coupling, short mean free path, regime the plasma represents a fluid (for a recent review, see \cite{Schafer:2009dj}), a number of recent studies initiated by \cite{son:2009tf} address the effects of triangle anomalies in hydrodynamics, e.g. \cite{Matsuo:2009xn,Sadofyev:2010is,Neiman:2010zi,Sadofyev:2010pr}. 

The central observation of the present paper is the following: the connection between the CME and CSE implies the existence of a new type of a collective excitation in the plasma. This excitation stems from the coupling between the density waves of electric and chiral charge. Let us illustrate this statement by a qualitative argument, to be followed by more rigorous derivations in sections \ref{mhd}, \ref{section3}, and \ref{section4}.  
Consider a local fluctuation of electric charge density; according to eq.(\ref{cse}) it will induce a local fluctuation of axial current. This fluctuation of axial current would in turn induce a local fluctuation of the axial chemical potential, and thus according to eq.(\ref{cme}) a fluctuation of electric current. The resulting 
fluctuation of electric charge density completes the cycle leading to the excitation that combines the density waves of electric and chiral charges; we will call it the "chiral magnetic wave" (CMW).

Apart from being interesting in its own right, the existence of CMW has important implications for the phenomenology of heavy ion collisions. The CME relies on the fluctuation of the axial charge density and so the net effect is expected to vanish when averaged over many events; one thus relies on measuring the fluctuations of charge asymmetries \cite{Kharzeev:2004ey,Voloshin:2004vk}. On the other hand, since the quark-gluon plasma produced in heavy ion collisions possesses non-zero value of the baryon chemical potential, the CSE can lead to a non-vanishing axial current even after the summation over events is performed. However unfortunately a direct detection of the axial current in heavy ion collisions is very challenging. The CMW should exist even in a neutral plasma and so can induce interesting observable effects in heavy ion collisions even after the sum over many events is performed; we will return to this topic in the Summary.

The paper is organized as follows. In section \ref{mhd} we provide a derivation of the CMW based on relativistic magnetohydrodynamics. In section \ref{section3} we consider the case of a strong magnetic field and perform a dimensional reduction; in this case the dynamics of CMW is described by the Sine-Gordon equation. In section \ref{section4} we describe the CMW at strong coupling using the holographic methods within the Sakai-Sugimoto model. When the electromagnetism is treated dynamically, the CMW mixes with the longitudinal charge wave in the plasma -- the plasmon. We consider the mixing of CMW with plasmons in section \ref{section5}. Finally, in the Summary we outline the main result of the paper and discuss the directions for future studies.

\section{Chiral magnetic wave in magnetohydrodynamics}\label{mhd}

Let us now proceed with the derivation sketched out in the introduction. We will see that there indeed 
exists a new gapless excitation in a deconfined QCD
plasma that propagates along the applied magnetic field; it arises as a dynamical consequence of the underlying triangle anomaly of chiral symmetry.
This new excitation is a long wavelength hydrodynamic mode with a dispersion relation that looks like that of sound waves,
\be
\omega=\mp v_\chi k - iD_L k^2 +\cdots\quad;
\ee
however, these propagating modes carry both electric and chiral charges. Since these modes would not exist if it were not for the applied magnetic field or the underlying triangle anomaly, we will call them "the chiral magnetic waves" (CMW). They give rise to several important new transport properties of hot QCD plasma, and affect its thermodynamics; we will further discuss this in section \ref{section5}. We note that CMW exists even if the background plasma is neutral under either baryonic or axial symmetry, which should make it a  generic phenomenon in relativistic plasmas.

For simplicity, let us consider single flavor ($N_F=1$) massless QCD with chiral symmetry $U(1)_L\times U(1)_R$, or equivalently $U(1)_V\times U(1)_A$ where $V(A)$ denotes vector(axial) respectively. The axial symmetry $U(1)_A$ suffers from both QCD anomaly with gluonic topological density and from the triangle anomaly of global chiral symmetry. The latter is in fact not harmful to the conservation of $U(1)_A$ as long as one does not elevate the global chiral symmetry to a gauged one, while the former indeed breaks the axial $U(1)_A$ symmetry by quantum fluctuations of topological density.

Our starting point is the anomalous generation of vector and axial currents along the applied magnetic field in the presence of axial (vector) chemical potential $\mu_A$ ($\mu_V$), as given by eqs. (\ref{cme}) and (\ref{cse}). We will now re-write these equations 
 in a more suggestive matrix form as
\be
\left(\begin{array}{c} \vec j_V \\ \vec j_A\end{array}\right) = {N_c\ e \vec B\over 2\pi^2} \left(\begin{array}{cc}
0 & 1 \\ 1 & 0
\end{array}\right) \left(\begin{array}{c}
\mu_V \\ \mu_A
\end{array}\right)\quad.
\ee
We are interested in small linearized fluctuations of the chiral currents $j_A$ and $j_V$ in the plasma; let us assume that this plasma is neutral, without any background charge density on average. We may then perform a linear expansion of the chemical potentials with respect to small charge densities $(j^0_V,j^0_A)$,
\be
\left(\begin{array}{c}
 \mu_V \\ \mu_A
\end{array}\right)=\left(\begin{array}{cc}
\partial\mu_V\over \partial j^0_V & \partial\mu_V\over \partial j^0_A \\
 \partial\mu_A\over \partial j^0_V  & \partial\mu_A\over \partial j^0_A 
\end{array}\right) \left(\begin{array}{c}
 j^0_V \\ j^0_A 
\end{array}\right) + {\cal O}\left(\left(j^0\right)^2\right) \equiv
\left(\begin{array}{cc}
\alpha_{VV}& \alpha_{VA}\\
 \alpha_{AV}  & \alpha_{AA} 
\end{array}\right) \left(\begin{array}{c}
 j^0_V \\ j^0_A 
\end{array}\right)+ {\cal O}\left(\left(j^0\right)^2\right)
\quad.
\ee
Remembering that 
\be
\mu_i = {\partial {\cal F}\over \partial j^0_i}\quad,\quad i=V,A
\ee
where $\cal F$ is the Helmholtz free energy, the $\alpha$'s appearing above are nothing but the susceptibility matrices 
of vector/axial charge densities,
\be
\alpha_{ij} = {\partial^2 {\cal F}\over \partial j^0_i \partial j^0_j}\quad.
\ee
Considering the parity $\cal{P}$ transformation $V\to - V$ and $A\to A$, one concludes that parity invariance of QCD implies that 
$\alpha_{VA}=\alpha_{AV}=0$ in the neutral plasma, $\mu_V=\mu_A=0$. Moreover, a simple large $N_c$ counting shows that
\be
\alpha_{VV}\sim \alpha_{AA} \sim {\cal O}\left(1\over N_c\right)\quad,
\ee
while their difference in a deconfined and chirally symmetric phase is subleading
\be
\alpha_{VV}- \alpha_{AA} \sim {\cal O}\left(1\over N_c^2\right)\quad;
\ee
we will confirm this within the holographic large $N_c$ Sakai-Sugimoto model in section \ref{section4}.
Independently of this, the vanishing of the difference $\alpha_{VV}-\alpha_{AA}$ can be taken as
a signal of chiral symmetry restoration. Therefore, we expect it to be a good approximation to let $\alpha_{VV}=\alpha_{AA}\equiv \alpha$ in the chirally symmetric phase; this leads us to 
\be
\left(\begin{array}{c} \vec j_V \\ \vec j_A\end{array}\right) = {N_c\ e \vec B \alpha \over 2\pi^2} \left(\begin{array}{cc}
0 & 1 \\ 1 & 0
\end{array}\right) \left(\begin{array}{c}
j^0_V \\ j^0_A
\end{array}\right)\quad.\label{semifinal}
\ee

It is natural to diagonalize the equation above by going to the chiral basis
\be
j^\mu_L\equiv {1\over 2}\left(j^\mu_V-j^\mu_A\right)\quad,\quad j^\mu_R\equiv
{1\over 2}\left(j^\mu_V+j^\mu_A\right)\quad.
\ee
In terms of chiral currents, our previous assumptions and the definition of $\alpha$'s are easily translated to
\be
\alpha={1\over 2}\left(\partial \mu_L\over \partial j^0_L \right)={1\over 2}\left(\partial^2 {\cal F}\over \partial j^0_L \partial j^0_L\right)=
{1\over 2}\left(\partial \mu_R \over \partial j^0_R\right)={1\over 2}\left(\partial^2 {\cal F}\over \partial j^0_R \partial j^0_R\right)\quad.
\ee
The (\ref{semifinal}) then leads to two decoupled relations
\be
\vec j_{L,R}= \mp \left(N_c e \vec B \alpha\over 2\pi^2\right) j^0_{L,R}\quad,\label{consti}
\ee
where one should keep in mind the definite sign in front of the right-hand side depending on the chirality of the currents.

One can view the above expression as the leading constitutive equation for the currents in the long wavelength derivative expansion of hydrodynamics.
Indeed, our starting point (\ref{cme},\ref{cse}) is strictly valid only when the variation of chemical potentials is sufficiently slow; for a finite frequency/momentum these expression gets modified resulting in frequency/momentum dependent chiral magnetic conductivity \cite{Kharzeev:2009pj,Yee:2009vw,Fukushima:2010vw,Fukushima:2009ft}. 
The equation (\ref{consti}) is the first leading term in the derivative expansion, while the next leading-order correction to the
chiral magnetic conductivity will be $\partial^2$ or $\omega^2\sim k^2$ in frequency/momentum space.
However, there is an important first-order derivative term in any constitutive equation of conserved current: a diffusion term  $-D \vec\nabla j^0$, with a diffusion constant $D$. In our case, we will be interested only in the waves propagating along the magnetic field direction which we call {\it longitudinal}; thus on general grounds, the constitutive relation including the next leading-order diffusion term reads as
\be
\vec j_{L,R}= \mp \left(N_c\ e \vec B \alpha\over 2\pi^2\right) j^0_{L,R} - D_L {\vec B (\vec B\cdot \vec\nabla)\over B^2} j^0_{L,R} +\cdots \quad,\label{consti2}
\ee
with a longitudinal diffusion constant $D_L$. Although we discuss only longitudinal dynamics in this paper,
it would also be interesting to study the transverse dynamics with the transverse diffusion constant $D_T$.

A similar constitutive equation was written previously by Son and Surowka \cite{son:2009tf}. There is however one point that will appear important for us: 
while Ref.\cite{son:2009tf} considers a weak magnetic field $B$ and treats it in the linear approximation, we are claiming that (\ref{consti}) and (\ref{consti2}) are valid for {\it arbitrary} strength of $eB$ non-perturbatively. 
This is equivalent to the validity of our starting point (\ref{cme}),(\ref{cse}) for arbitrarily large $eB$, which is not at all trivial and is a consequence of the absence of  corrections to the axial anomaly. 
It is also important to note that although (\ref{cme}),(\ref{cse}) look linear in $eB$, this linearity is only apparent.
Given a fixed density $j^0_{V,A}$, the chemical potentials $\mu_{V,A}$ in general may well depend
on the dynamics of underlying microscopic theory, such as coupling constants, temperature, as well as magnetic field $eB$ {\it non-linearly}, so that
the currents in (\ref{cme}),(\ref{cse}) can in fact be very non-linear in these parameters, see e.g. \cite{Gorbar:2009bm,Fukushima:2010zz,Gorbar:2010kc}. The statement of (\ref{cme}),(\ref{cse}) is that
these dependencies can be absorbed into the chemical potentials $\mu_{V,A}$. 
Therefore, one expects that $\alpha$ and $D_L$ are in general non-linear functions of $eB$, temperature $T$, etc. In particular, they would also
depend on the coupling constant, so that it is meaningful to study them in the strong coupling regime via holographic QCD as we do in our section \ref{section4}. 
Let us mention Ref.\cite{hong} that presents a diagrammatic proof of (\ref{cme}),(\ref{cse}) perturbatively
in the coupling constant, and Ref.\cite{Rebhan:2009vc} that proved this relation in the Sakai-Sugimoto model using the  two-derivative approximation. We will present a strong coupling proof with full DBI action of holographic QCD in section \ref{section4} with arbitrary strength of $eB$, which presumably includes non-linear effects of derivatives as well. Therefore we expect that the equations (\ref{cme}),(\ref{cse}) and hence (\ref{consti2}) hold universally.

Our next step is to combine (\ref{consti2}) with the conservation law $\partial_\mu j^\mu_{L,R}=0$. We take $\vec B=B \hat x^1$ and consider only longitudinal gradient $\partial_1$, which results in
\be
\left(\partial_0 \mp{N_c e B \alpha \over 2\pi^2} \partial_1 -D_L \partial^2_1 \right) j^0_{L,R}=0\quad.
\ee 
This describes a directional wave, or {\it chiral wave}, of charge densities whose direction of motion is correlated with its chirality.
The velocity is given by
\be\label{velocity}
v_\chi={N_c e B \alpha \over 2\pi^2}= {N_c e B\over 4\pi^2} \left(\partial \mu_L\over \partial j^0_L\right)={N_c e B\over 4\pi^2} \left(\partial\mu_R\over \partial j^0_R\right)\quad.
\ee
As we discussed in the previous paragraph, one expects that $v_\chi$ and $D_L$ are non-trivial functions of $eB$, $T$, and the coupling constant, so
they are interesting dynamical quantities to compute in any model.
In frequency/momentum space, the above equation takes the form
\be\label{dispersion}
\omega = \mp v_\chi k - i D_L k^2 +\cdots \quad,
\ee
as a hydrodynamic dispersion relation. Our main observation is the new first term in the dispersion relation which makes the mode propagating
instead of simply diffusing. It exists only if 1) triangle anomaly exists and 2) there is a background magnetic field. 
We stress that this new chiral mode of electric and chiral charge transport is present even if the plasma is neutral on average.

\section{Chiral magnetic wave in strong magnetic field:\break dimensional reduction to $(1+1)$ Sine-Gordon problem \label{section3}}

The expression (\ref{velocity}) for the velocity $v_\chi$ of the Chiral Magnetic Wave (CMW) shows that  $v_\chi \sim eB$. 
What happens when $eB$ becomes large? We will now show that in the limit $eB \to \infty$ the velocity $v_\chi$ stays finite and reaches the velocity of light. To understand this, let us first examine the spectrum of charged fermions in magnetic field; for massless fermions, the energies of Landau levels are given by
\be
E_n = \sqrt{2 e B (n + \frac{1}{2} - s_z) + p_z^2};
\ee
we assume that magnetic field ${\vec B}$ is directed along the $z$ axis.
 Note that the lowest Landau level (LLL) with zero energy (at vanishing momentum $p_z$) is not spin-degenerate, and all excited Landau levels are -- therefore the net chirality $s_z \cdot p_z$ is carried only by the LLLs. 

When magnetic field is large compared to the temperature $T$, $eB \gg T^2$, the fermions stay "frozen" in the LLL and the transverse and longitudinal (along $\vec z$) dynamics are independent. In this case one can perform a dimensional reduction to the $(1+1)$ dimensional theory where the only allowed direction of motion is along the magnetic field.  The transverse density of states is given by $eB/(2\pi)$, the density of LLs in the transverse plane. The longitudinal phase space density for a Fermi-momentum  is simply $p_F/(2 \pi)$. Therefore for the density of $N_c$ left-handed fermions $j^0_L$ with $p_F = \mu_L$
(for massless particles the Fermi-momentum and the chemical potential $\mu$ are equal) is 
\be 
j^0_L =  N_c\  \frac{eB}{2\pi}\ \frac{\mu_L}{2\pi},
\ee
and the derivative is given by
\be
\frac{\partial \mu_L}{\partial j^0_L} = \frac{1}{N_c}\ \frac{4 \pi^2}{e B}.
\ee
Substituting this expression into (\ref{velocity}), we obtain
\be 
v_\chi=1;
\ee
therefore in the limit of strong magnetic field $eB \gg T^2$ the CMW indeed propagates with the velocity of light. 

We can deduce a more detailed information about the dynamics of the CMW in strong magnetic field by making use of bosonization 
procedure. As is well known, bosonization approach is very powerful in the studies of $(1+1)$ dimensional systems \cite{Coleman:1974bu,Mandelstam:1975hb}. This is easy to understand as in one spatial dimension the produced fermion and an anti-fermion never separate and propagate together as a composite bosonic excitation -- even if they do not interact at all! 

The conservation of vector current $\partial_\mu j^\mu_V = 0$ can be ensured by introducing a boson field $\varphi$ and 
choosing 
\be
j^\mu_V =  \frac{1}{\sqrt{\pi}} \epsilon^{\mu\nu} \partial_\nu \varphi;
\ee
this way the vector current is always conserved independently of the equations of motion. The corresponding choice for the axial current is 
\be
j^\mu_A =   \epsilon^{\mu\nu} j_\nu^V = \frac{1}{\sqrt{\pi}} \partial^\mu \varphi.
\ee
For zero quark mass and in the absence of background electric field, the axial current should be conserved:
\be\label{waveeq}
\partial_\mu j^\mu_A =  \Box \varphi = 0.
\ee
Therefore the conservation of axial current leads to the wave equation for the bosonic excitation $\varphi$. 
The variation of $\varphi$ in space and time causes variations of both charge and chiral densities; therefore it is natural to identify the wave defined by (\ref{waveeq}) with the CMW. Since (for massless quarks) there is no mass term in (\ref{waveeq}), this CMW propagates with the velocity of light.

It is convenient to split the field $\varphi$ into left- and right-moving components:
\be
\varphi(x, t) = \varphi_L (x- vt)  + \varphi_R (x+ vt);
\ee
in terms of these fields the original fermion fields are 
\be\label{fermi_bose}
\Psi_R(x- vt) = \frac{1}{\sqrt{2\pi}}\ e^{- i \sqrt{4 \pi}  \varphi_R (x- vt)}; \ \ \ 
\Psi_L(x+ vt) = \frac{1}{\sqrt{2\pi}}\ e^{ i \sqrt{4 \pi}  \varphi_L (x+ vt)}.
\ee
Upon quantization of the boson fields, the fermions (\ref{fermi_bose}) represent coherent states of $\varphi_R, \varphi_L$. Viceversa, these boson fields 
describe the collective electric and chiral charge density fluctuations of the underlying fermion quark fields. 

\begin{figure}[t]
	\centering
	\includegraphics[width=10cm]{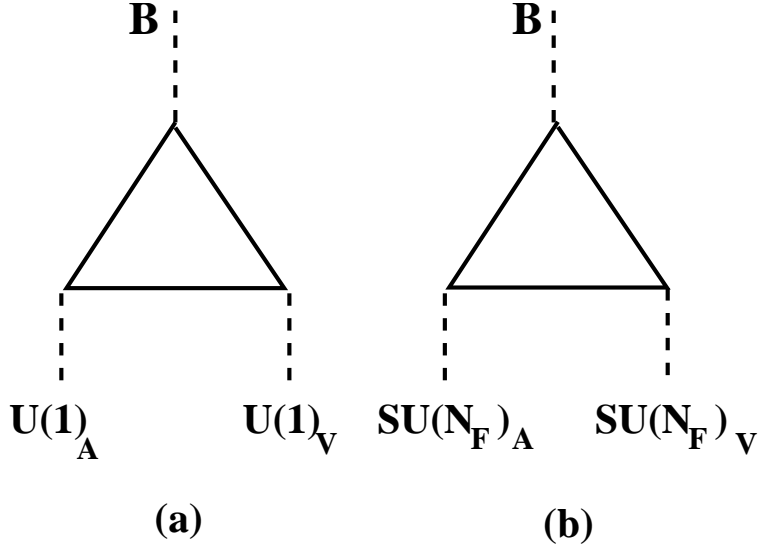}
		\caption{Anomalous triangle diagrams of chiral magnetic waves for (a) abelian and (b) non-abelian flavor symmetries in the presence of
external magnetic field $B$. The quark lines are dressed propagators including the background magnetic field. \label{figna}}
\end{figure}

The quark mass term in the original Lagrangian can be re-written with the help of (\ref{fermi_bose}) in the following form: 
\be
M \bar{\Psi} \Psi = M(\bar{\Psi}_R \Psi_L + \bar{\Psi}_L \Psi_R) = \frac{M}{\pi} \cos \left( \sqrt{4 \pi} \varphi \right),
\ee
and leads to the interaction term for bosons; the bosonic Lagrangian thus describes the Sine-Gordon theory. This correspondence between the fermionic and bosonic description has been discovered by Coleman \cite{Coleman:1974bu} and Mandelstam \cite{Mandelstam:1975hb}. The kinks of the Sine-Gordon theory describe the fermions, and the fluctuations of the boson field - collective fermion--anti-fermion excitations. 
Coleman's theorem \cite{Coleman:1973ci} forbids spontaneous breaking of a continuous symmetry  in $(1+1)$ dimensions. 
The underlying reason for this no-go theorem is the presence of strong infra-red fluctuations that dominate two-point functions and destroy the long-range order in two dimensions. This means that the massless (in the limit of massless quarks) boson $\varphi$ is not a Goldstone boson\footnote{There exists however a way around Coleman's theorem uncovered by Witten \cite{Witten:1978qu}: for $N$-component field at $N\to\infty$, when the number of degrees of freedom diverges at each point, the two-point functions can exhibit long-range order and the Goldstone phenomenon can still be realized.}. This is consistent with our interpretation of the field $\varphi$ as of a collective density wave of electric and chiral charges. At finite density of baryon or chiral charge, this density wave propagates on top of a "chiral spiral" \cite{Schon:2000qy} -- the winding configuration of the background field $\bar{\varphi}$ causing the chiral magnetic effect in $(1+1)$ dimensional description \cite{Basar:2010zd,holog_spiral}.

One can also generalize the above to non-abelian version of chiral magnetic waves. As Figure.\ref{figna} illustrates, essentially the same type
of triangle diagrams for mixed $U(1)_V\times SU(N_F)_V\times SU(N_F)_A$ would result in 
\be
\left(\begin{array}{c} \vec j_V^a \\ \vec j_A^a\end{array}\right) = {N_c e \vec B \alpha \over 2\pi^2} \left(\begin{array}{cc}
0 & 1 \\ 1 & 0
\end{array}\right) \left(\begin{array}{c}
j^{0a}_V \\ j^{0a}_A
\end{array}\right)\quad,\label{nonabelian}
\ee
for non-abelian $SU(N_F)$ components of currents $j^\mu_{V,A}=j^{\mu a}_{V,A} t^a$ with ${\rm tr}_F(t^a t^b)={1\over 2}\delta^{ab}$. 
One then gets to the same conclusion on the emergence of the chiral/directional CMWs for each non-abelian component of $SU(N_F)$.
Upon the 1+1 dimensional reduction with strong magnetic field, these non-abelian chiral magnetic waves should be described
by non-abelian bosonization \cite{Witten:1983ar} of $SU(N_F)$ symmetry. More precisely, the theory of $N_F$ Dirac fermions in fundamental representation of $SU(N_c)$ in bosonized description can be represented as 
\be
{\cal L}=N_c {\cal L}\left(SU(N_F)\right)+{\cal L}\left(U(1)\right)+N_F{\cal L}\left(SU(N_c)\right)\quad,
\ee
where ${\cal L}(G)$ represents the Wess-Zumino-Witten model of group $G$ with level 1 \cite{Witten:1983ar}.
After integrating over QCD $SU(N_c)$ dynamics, one is left with the first two pieces as a low energy effective theory \cite{Gonzales:1984zw,Gepner:1984au,Affleck:1985wa,Date:1986xe}. The $U(1)$ part
is what we have discussed above, while the non-abelian $SU(N_F)$ part describes the non-abelian chiral magnetic wave as a propagating
group field $g(x)\in SU(N_F)$.

\section{Chiral magnetic wave in holographic QCD \label{section4}}

In this section, we intend to study chiral magnetic wave at strong coupling
in the framework of holographic QCD, particularly using the model by Sakai and Sugimoto \cite{Sakai:2004cn}. The Sakai-Sugimoto model is a right place to look for the phenomenon
because it includes the relevant triangle anomalies of QCD chiral symmetry  
in terms of 5D Chern-Simons terms and  the existence of chiral magnetic wave 
is robust as long as the right anomalies exist. 
What is non-trivial will be the details of dispersion relations such as wave velocity and diffusion constant, which do depend on the strong dynamics and the stength of the applied magnetic field.

\begin{figure}[t]
	\centering
	\includegraphics[width=10cm]{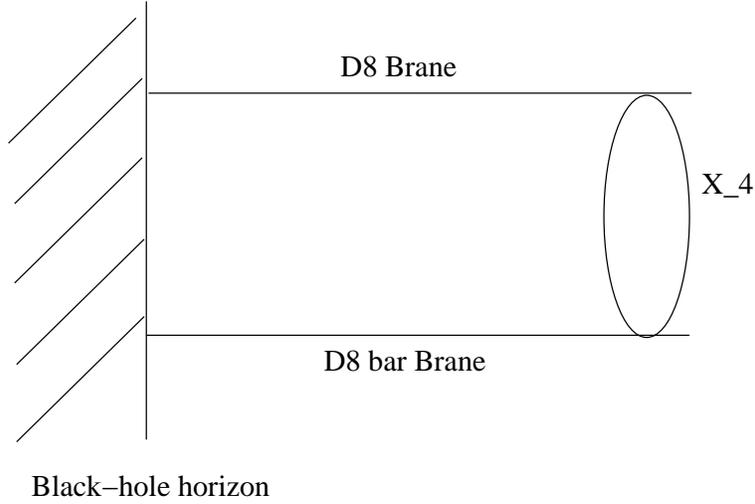}
		\caption{A schematic picture of $D8-\overline{D8}$ branes in deconfined phase of the Sakai-Sugimoto model. \label{fig0}}
\end{figure}
The Sakai-Sugimoto model in deconfined phase 
consists of two separated $D8$ and $\overline{D8}$ branes 
that touch the black hole horizon of the background geometry as shown in Figure \ref{fig0}. This geometric separation of two probe branes correctly indicates that interactions between left-handed quarks and right-handed quarks are sub-leading in large $N_c$ expansion, when there is no chiral condensate in the deconfined phase.  
Because of this decoupling, one can study dynamics of each $D8$ probe brane
independently at least in leading order approximation.
The background geometry is a warped product of a 5D black hole and $S^1\times S^4$ where $D8/\overline{D8}$ branes are separated along $S^1$. They wrap all other dimensions except $S^1$ to make up 9 dimensional world-volume on which the action is given by a Dirac-Born-Infeld action plus a Chern-Simons term;
\be
S_{D8/\overline{D8}} = -\mu_8 \int d^9 \xi \, e^{-\phi}\,\sqrt{{\rm det}\left(g^*+2\pi l_s^2 F\right)}
\,\,{\mp}\,\, {\mu_8\left(2\pi l_s^2\right)^3 \over 3!}\int \,F_4^{RR}\wedge A\wedge F\wedge F\quad,\label{8-act}
\ee
with $\mu_p=(2\pi)^{-p}l_s^{-(p+1)}$.
 After integrating over $S^4$, one arrives at an effective 5 dimensional world-volume action
of $D8/\overline{D8}$ branes embedded in the 5 dimensional black hole space-time
of the metric
\be
ds^2_{5D}=\left(U\over R\right)^{3\over 2}\left(-f(U)dt^2+\sum_{i=1}^3\left(dx^i\right)^2 \right)+2dUdt\quad,\quad f(U)=1-\left(U_T\over U\right)^3\quad,
\ee
where $U$ is the holographic direction whose UV(IR) boundary is sitting at $U\to\infty$($U=U_T$), and the metric is written in Eddington-Finkelstein coordinate.
Note that Eddington-Finkelstein coordinate is suitable to describe future event horizon in a smooth manner, and the in-going boundary conditions of perturbed modes at the future horizon reduce to simple regularity in this coordinate.
The temperature $T$ of the dual 4 dimensional QCD plasma is related to the parameters in the above by
\be
T={3\over 4\pi}\left(U_T\over R^3\right)^{1\over 2}\quad,\quad R^3=\pi g_s N_c l_s^3\quad.
\ee

The resulting 5 dimensional world-volume action of the probe brane
is
\bear
S_{D8/\overline{D8}}&=&-C R^{9\over 4}\int d^4x dU\, U^{1\over 4}\sqrt{{\rm det}\left(g^*_{5D}+2\pi l_s^2 F\right)}\nonumber\\ &\mp& {N_c\over 96\pi^2}\int d^4x dU \, \epsilon^{MNPQR} A_M F_{NP} F_{QR}\quad,\quad C={N_c^{1\over 2}\over 3\cdot 2^5 \pi^{11\over 2} g_s^{1\over 2} l_s^{15\over 2}}\quad,\label{5daction}
\eear
where the sign of the last Chern-Simons term depends on the chirality of the probe $D8$/$\overline{D8}$ branes. The $\epsilon$-symbol above is numerical one. 
It is straightforward to write down the equation of motion of the world-volume gauge field $A_M$ from the above action, which reads as
\bear
C R^{9\over 4}\pi l_s^2\,\partial_N\left[U^{1\over 4}\sqrt{{\rm det}\left(g^*+2\pi l_s^2 F\right)} \left(\left(g^*+2\pi l_s^2 F\right)^{-1}\right)^{[MN]}\right]\
\mp {N_c\over 32\pi^2}\epsilon^{MNPQR}F_{NP} F_{QR} =0,\nonumber\\\label{eom}
\eear
where $[MN]=MN-NM$ is anti-symmetrization.
In deriving the above as well as for later convenience, it is useful to be reminded of the following expansion
\be
\sqrt{{\rm det}\left(1+\delta A\right)}=1+{1\over 2}{\rm tr}\left(\delta A\right)+{1\over 8}\left[{\rm tr}\left(\delta A\right)\right]^2-{1\over 4}{\rm tr}\left(\left(\delta A\right)^2\right)+{\cal O}\left(\left(\delta A\right)^3\right)\quad.\label{detexp}
\ee

 For our purpose of studying chiral magnetic wave, we first need to construct a background solution of having a constant magnetic field of {\it arbitrary strength} pointing, say, $x^3$ direction.
It is in fact easy to show that the trivially constant $F_{12}\equiv B$ solves both Bianchi identity and the equation of motion (\ref{eom}), so the
{\it holographic} background configuration of a constant $B$ will be our starting point of looking for a signal of chiral magnetic wave.
In making contact between our $D8$/$\overline{D8}$ branes and the QCD chiral symmetry, we work in the case of $N_F=1$ for simplicity, neglecting $U(1)_A$-anomaly induced by gluons as a leading large $N_c$ approximation. This is appropriate as long as we focus on triagle anomalies which are encoded in the 5D Chern-Simons terms in holographic QCD. Therefore, chiral symmetry is $U(1)_L\times U(1)_R$, and $D8$($\overline{D8}$)-brane dynamics captures $U(1)_L$($U(1)_R$) chiral dynamics of QCD holographically. The electromagnetism lies in the diagonal combination of $U(1)_L$ and $U(1)_R$, while axial symmetry is the other orthogonal combination, so one has a dictionary
\be
eA_{EM}={1\over 2}\left(A_L+A_R\right)\quad,\quad A_a={1\over 2}\left(-A_L+A_R\right)\quad,
\ee
where $A_L$($A_R$) is the external potential that couples to the chiral current $J_L$($J_R$), and $e$ is the electromagnetic coupling constant. Therefore, having a constant electromagnetic $B$ means 
having a constant $F_{12}$ on each $D8$ and $\overline{D8}$ brane with
\be
F_{12}^{D8}=F_{12}^{\overline{D8}}= eB\quad,
\ee
and we will assume this in the following.

Our next task is to expand linearly around this background to study hydrodynamics of {\it longitudinal} charge/current fluctuations.
As we discuss in previous sections, these longitudinal charge/current fluctuations would have had leading diffusive dispersion relation $\omega\sim -iD k^2$ in the absence of triangle anomalies.  Chiral magnetic wave is an anomaly-induced modification of this into a leading propagating dispersion relation 
\be
\omega\sim \mp v_\chi k -iD_L k^2 +\cdots\quad,\label{dispersion}
\ee
where the velocity $v_\chi$ as well as the {\it longitudinal} diffusion constant $D_L$ are expected
to depend on strong coupling dynamics and the magnitude of the magnetic field $B$. The above dispersion relation due to triangle anomalies is our main objective we will be heading to in this section.  Especially, $v_\chi$ as a function of $B$ will be a new physical quantity that we compute in the framework of holographic QCD. Because $v_\chi$ should vanish in the absence of anomalies and the magnetic field, it will be an odd function of the product of Chern-Simons coefficient and the magnetic field $B$. We will confirm this expectation later.
 We also emphasize that the sign of the leading term in (\ref{dispersion}) is fixed by the sign of Chern-Simons term, that is, by the chirality one is looking at, so left-handed charge/current fluctuations would have a definite propagating direction with respect to the magnetic field direction, while
the right-handed ones would propagate in the opposite way. Therefore, one can achieve {\it chirality separation} in this way as a dynamical consequence of triangle anomalies of QCD. Recall that no charge in the background, either axial or baryonic, needs to be present for the effect to take place, and only magnetic field applied to a deconfined plasma with an underlying triangle anomaly is sufficient to have the phenomenon.

Studying linearized fluctuations from the constant $F_{12}=eB$ background requires expanding the 5D action (\ref{5daction}) quadratically in terms of relavant fluctuation fields, and it is tedious but straightforward to do it using the formula (\ref{detexp}). As we are interested in longitudinal charge/current fluctuations, it is sufficient to consider  $(\delta F_{tU},\delta F_{3U},\delta F_{t3})$
fluctuations of ``helicity'' 0 only. Because of residual $SO(2)$ rotation symmetry of the constant $F_{12}$ background, other non-zero helicity modes simply decouple
from the above modes at linearized level of equations of motion.
After a sizable amount of computation, one arrives at
\bear
S_{D8/\overline{D8}}^{(2)}&=&\int d^4x dU \,{1\over 2} \left[A(U)\left(\delta F_{tU}\right)^2-B(U) \left(\delta F_{3U}\right)^2 +2 C(U)\left(\delta F_{t3}\right)\left(\delta F_{3U}\right)\right] \nonumber\\
&\mp& {N_c eB\over 8\pi^2}\int d^4x dU \left[ \delta A_U \delta F_{t3}
-\delta A_3 \delta F_{tU}+\delta A_t \delta F_{3U}\right]\quad,\label{quadexp}
\eear 
where the three functions that appear in the coefficients are given by
\bear
A(U)&=& C\left(2\pi l_s^2\right)^2 U \left[U^3 +R^3 \left(2\pi l_s^2 eB\right)^2\right]^{1\over 2}\quad,\nonumber\\
B(U)&=&C\left(2\pi l_s^2\right)^2 U f(U)\left[U^3 +R^3 \left(2\pi l_s^2 eB\right)^2\right]^{1\over 2}\quad,\nonumber\\
C(U)&=&    C\left(2\pi l_s^2\right)^2 U \left(R\over U\right)^{3\over 2}\left[U^3 +R^3 \left(2\pi l_s^2 eB\right)^2\right]^{1\over 2} \quad.\label{coeff}
\eear
The second line in (\ref{quadexp}) is from the Chern-Simons term that represents triangle anomaly of QCD holographically. It is easy to keep track of that by a combination $N_c e B$ which is the  product of anomaly coefficient and the magnetic field.
From the above quadratic expansion of the action, one easily writes down the linearized equations of motion as
\bear
\partial_U\left(A(U)\delta F_{tU}\right)+C(U)\left(\partial_3 \delta F_{3U}\right)\mp {N_c eB\over 4\pi^2} \delta F_{3U} &=& 0\quad,\nonumber\\
\partial_U\left(B(U)\delta F_{3U}\right)+C(U)\left(\partial_t \delta F_{3U}\right)-\partial_U\left(C(U)\delta F_{t3}\right) \mp {N_c eB\over 4\pi^2} \delta F_{tU} &=& 0\quad,\nonumber\\
A(U)\left(\partial_t \delta F_{tU}\right)-B(U)\left(\partial_3 \delta F_{3U}\right)+C(U)\left(\partial_3 \delta F_{t3}\right) \pm {N_ceB\over 4\pi^2}\delta F_{t3}&=&0\quad.\label{lineom}
\eear 
To proceed to hydrodynamic analysis from the above, one invokes low frequency/momentum expansion in solving (\ref{lineom}) order by order with right boundary conditions.
As mentioned before, simple regualarity at the future horizon $U=U_T$ in Eddington-Finkelstein coordinate is equivalent to implementing in-coming boundary conditions, and one should also impose normalizability on the modes at the UV boundary $U\to\infty$. Near $U\to\infty$, (\ref{lineom}) gives two asymptotic behaviors, either ${\cal O}(1)$ or ${\cal O}(U^{-{3\over 2}})$, and the normalizability picks up the latter only.  
Assuming the frequency/momentum factor $e^{-i\omega t+i k x^3}$ or equivalently replacing $\partial_t=-i\omega$ and $\partial_3=ik$, and working in a gauge $A_U=0$, the equations of motion become
\bear
\partial_U\left[A(U)\left(\partial_U \delta A_t\right)\right)]+ik C(U)\left(\partial_U\delta A_3\right)\mp {N_c eB\over 4\pi^2}\left(\partial_U\delta A_3\right)=0,&&\nonumber\\
\partial_U\left[B(U)\left(\partial_U\delta A_3\right)\right]-i\omega
C(U)\left(\partial_U\delta A_3\right)-\partial_U\left[C(U)
\left(i\omega \delta A_3 +ik\delta A_t\right)\right]\mp {N_c eB\over 4\pi^2}\left(\partial_U\delta A_t\right)= 0,&&\nonumber\\
-i\omega A(U)\left(\partial_U\delta A_t\right)-ik B(U)\left(\partial_U\delta A_3\right)+ik C(U)\left(i\omega \delta A_3 +ik\delta A_t\right)\pm
{N_ceB\over 4\pi^2}\left(i\omega \delta A_3 +ik \delta A_t\right)=0,&&\nonumber\\\label{hydroeom}
\eear
which should be solved in perturbative expansion of $(\omega,k)$.
What one expects is that imposing boundary conditions restricts the solution space such that $\omega$ in hydrodynamic expansion is determined once $k$ is given, and the relation
$\omega=\omega(k)$ is called the {\it dispersion relation}.

There are several methods in literature to solve similar kinds of 
equations in hydrodynamic expansion \cite{Policastro:2002se}, but we will follow our own method
which seems most convenient to us. 
We first assume that $\omega(k)$ is an analytic power series in $k$, which is expected based on hydrodynamics,
\be
\omega(k) =\sum_{n\ge 1}a_n k^n= v_\chi k -iD_L k^2 +\cdots\quad.\label{omegaexp}
\ee
Inserting this to (\ref{hydroeom}), then one can take $k$ as the only expansion parameter in solving (\ref{hydroeom}) systematically, along which $a_n$ should also be determined order by order. Because (\ref{hydroeom}) is linear in $(\delta A_t,\delta A_3)$ one can always rescale them so that they start their $k$-expansion as
\be
(\delta A_t,\delta A_3)=\sum_{n\ge 0} (A_t^{(n)},A_3^{(n)}) k^n =(A_t^{(0)},A_3^{(0)})+(A_t^{(1)},A_3^{(1)})k+\cdots\quad,\label{Aexp}
\ee
where $(A_t^{(0)},A_3^{(0)})$ cannot vanish simultaneously by definition.
It is straightforward to insert (\ref{omegaexp}) and (\ref{Aexp}) into (\ref{hydroeom}), and solve order by order in $k$.

At ${\cal O}(k^0)$, one gets the equations
\bear
\partial_U\left[A(U)\left(\partial_U A_t^{(0)}\right)\right]\mp {N_c eB\over 4\pi^2}\left(\partial_UA_3^{(0)}\right)=0\quad,&&\label{zero1}\\
\partial_U\left[B(U)\left(\partial_U A_3^{(0)}\right)\right]\mp {N_c eB\over 4\pi^2}\left(\partial_U A_t^{(0)}\right)=0\quad,&&\label{zero2}\\
-iv_\chi A(U)\left(\partial_U A_t^{(0)}\right)-iB(U)\left(\partial_U A_3^{(0)}\right)\pm {N_c eB\over 4\pi^2}\left(i v_\chi A_3^{(0)}+i A_t^{(0)}\right)=0\quad.\label{zero3}
\eear
Integrating the first two equations (\ref{zero1}) and (\ref{zero2}) gives us
\bear
A(U)\left(\partial_U A_t^{(0)}\right)\mp {N_c eB\over 4\pi^2} A_3^{(0)}&=& C_1\quad,\label{C1}\\
B(U)\left(\partial_U A_3^{(0)}\right)\mp {N_c eB\over 4\pi^2}A_t^{(0)}&=& C_2\quad,\label{C2}
\eear
with two integration constants, while the last equation (\ref{zero3}) simply becomes
\be
-i v_\chi C_1 -i C_2 =0\quad \Longrightarrow  v_\chi=-{C_2\over C_1}\quad,
\ee
so that $v_\chi$ will be determined once $C_{1,2}$ (or more precisely their ratio) are fixed by imposing relevant boundary conditions.
Considering (\ref{C1}) at $U\to\infty$, one first fixes 
\be
C_1=\lim_{U\to\infty}A(U)\left(\partial_U A_t^{(0)}\right)\quad,
\ee
and solving $A_3^{(0)}$ gives
\be
A_3^{(0)}=\pm \left(4\pi^2\over N_c eB\right)\left(A(U)\left(\partial_U A_t^{(0)}\right)-C_1\right)\quad.
\ee
Inserting this into (\ref{C2}), one gets a second order differential equation for $A_t^{(0)}$,
\be
B(U)\partial_U\left(A(U)\left(\partial_U A_t^{(0)}\right)\right)
-\left(N_c eB\over 4\pi^2\right)^2 A_t^{(0)} =\pm C_2 \left(N_c eB\over 4\pi^2\right)\quad.
\ee
Because $B(U_T)=0$ at the horizon, the regularity boundary condition imples that
\be
C_2=\mp \left(N_c eB \over 4\pi^2\right)A_t^{(0)}(U_T)\quad,
\ee
and one can write the solution for $A_t^{(0)}$ as
\be
A_t^{(0)}=\tilde A +A_t^{(0)}(U_T)\quad,\label{at0}
\ee
where $\tilde A$ satisfies
\be
B(U)\partial_U\left(A(U)\left(\partial_U \tilde A \right)\right)
-\left(N_c eB\over 4\pi^2\right)^2\tilde A =0\quad,\label{tildeA}
\ee
with the boundary condition $\tilde A(U_T)=0$ at the horizon. 
This uniquely determines $\tilde A$ up to rescaling.
Note that $A_t^{(0)}(U_T)$ is free up to this point, and
the final boundary condition we need to impose is to demand vanishing $A_t^{(0)}$ at $U\to\infty$, and from (\ref{at0}) this fixes  $A_t^{(0)}(U_T)$ as
\be
A_t^{(0)}(U_T)=-\lim_{U\to\infty} \tilde A(U)\quad,
\ee
so that $C_2$ is finally 
\be
C_2=\pm \left(N_c eB \over 4\pi^2\right)\lim_{U\to\infty} \tilde A(U)\quad.
\ee
Observe that $C_1$ is also given by $\tilde A$ as
\be
C_1=\lim_{U\to\infty}A(U)\left(\partial_U \tilde A\right)\quad.
\ee
Therefore, the complete solution at ${\cal O}(k^0)$ with the right boundary conditions can be
written solely in terms of $\tilde A$ as above, and it is unique up to overall
rescaling. Especially $v_\chi$ is well-posed and given by
\be
v_\chi = -{C_2\over C_1}=\mp\left(N_c eB\over 4\pi^2\right)\lim_{U\to\infty}\left(\tilde A(U)\over A(U)\left(\partial_U \tilde A\right)\right)\quad.\label{vchifinal}
\ee
As expected, $v_\chi$ is proportional to the anomaly coefficient, and its sign depends on the chirality and the chiral magnetic wave is uni-directional.

Discussions in the previous sections independently argue that $v_\chi$ should be given by
\be
v_\chi = \mp \left(N_c eB\over 4\pi^2\right)\left(\partial \mu\over \partial j^0\right)_{j^0=0}\quad,\label{holoproof}
\ee
where $(\mu,j^0)$ are chemical potential and charge density for either $U(1)_L$ or $U(1)_R$. Indeed we can confirm this expectation in our final formula
(\ref{vchifinal}) which is fully non-linear in $eB$, so that one can consider this as a strong coupling proof of the relation.
First note that to compute $\left(\partial \mu\over \partial j^0\right)_{j^0=0}$ one only needs a linear perturbation of $\mu$ or $j^0$ to the system of our background $B$ field, and the relevant equations of motion for them are precisely
given by our previous one (\ref{lineom}) with additional assumption of space-time homogeneity $\partial_t=\partial_3=0$. Then the equations simplify exactly to 
the previous (\ref{C1}) and (\ref{C2}) with suitable boundary conditions.
In the second equation considering the horizon point $U=U_T$, we have $B(U_T)=0$ and we demand that $A_t$ vanishes at the horizon, so that $C_2=0$.
Combining the two equations after removing $A_3$, one easily arrives at that $A_t$ satisfies the same equation (\ref{tildeA}) that $\tilde A$ satisfies, and moreover they share the same boundary condition at the horizon $U=U_T$, so that
they are in fact the same object $A_t=\tilde A$. On the other hand, gauge/gravity dictionary
tells us that up to linear order, 
\be
\mu=\lim_{U\to\infty}A_t(U)\quad,\quad j^0=\lim_{U\to\infty} A(U)\left(\partial_U A_t\right)\quad,
\ee
so that one has
\be
\left(\partial \mu\over \partial j^0\right)_{j^0=0} =  \lim_{U\to\infty}\left(\tilde A(U)\over A(U)\left(\partial_U \tilde A\right)\right)\quad,
\ee
which proves the relation (\ref{holoproof}) at strong coupling fully non-linearly in $eB$. As the validity of (\ref{holoproof}) is equivalent to
(\ref{cme}),(\ref{cse}), this constitutes a holographic proof of (\ref{cme}),(\ref{cse}).

\begin{figure}[t]
	\centering
	\includegraphics[width=10cm]{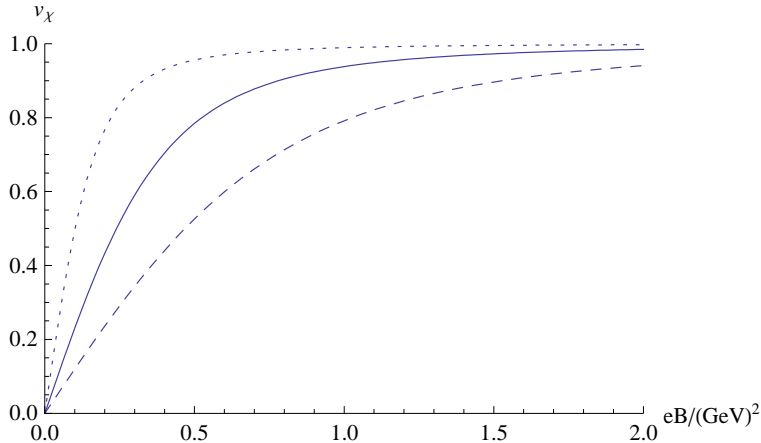}
		\caption{Numerical result for $v_\chi$ in the Sakai-Sugimoto model with $T=150$ MeV(dotted), $T=200$ MeV(plain), and $T=250$ MeV(dashed). \label{fig1}}
\end{figure}
It is interesting to see how $v_\chi$ depends on the magnitude of the magnetic field $eB$ in a non-linear way because $(A(U),B(U))$ contain $eB$ as in (\ref{coeff}), although the necessary analysis inevitably involves numerical study.
To perform numeric analysis, we have to specify parameters of the model.
First of all, one can always put $2\pi l_s^2\equiv 1$ for simplicity because this factor will eventually cancel out in any well-defined field theory observables. One can easily check that this is the case for $v_\chi$ as well.
By fitting to the observed $\rho$-meson mass and the pion decay constant, Sakai-Sugimoto fixed the parameters as
\be
N_c=3\quad,\quad g_{YM}^2 N_c \sim 17\quad,\quad M_{KK}\sim 0.94\,{\rm GeV}\quad,
\ee
where $g_{YM}^2$ and $M_{KK}$ are related to $g_s$ by
\be
g_{YM}^2= 2\pi l_s M_{KK} g_s\quad.
\ee
Recall that what matters for us is simply the parameters $R^3$ and $C$, and
in terms of the above parameters, one has
\be
C\sim0.0211\quad,\quad R^3 \sim 1.44\quad.
\ee
For the temperature, we take $T=(150,200,250)$ MeV as an illustrative purpose. Note that this model has deconfinement phase transition at $T_c={M_{KK}\over 2\pi}\sim 150$ MeV \cite{Aharony:2006da}. 
We plot our numeric result of $v_\chi$ as a function of $eB$ in Figure \ref{fig1}.

Limited analytic results for $v_\chi$ are available for two extreme regions, either $eB\to 0$ or $eB\to\infty$. For this purpose as well as an easier numerical analysis, it is convenient to consider the combination
\be
V(U)\equiv {\tilde A(U)\over A(U)\left(\partial_U \tilde A\right)}\quad,
\ee
in terms of which the equation (\ref{tildeA}) becomes a simple first order differential equation
\be
\partial_U V(U)={1\over A(U)}-\left(N_c eB\over 4\pi^2\right)^2 {1\over B(U)}V^2(U)\quad,
\ee
with a boundary condition $V(U_T)=0$. From (\ref{vchifinal}), $v_\chi$ is then simply given by
\be
v_\chi=\mp\left(N_c eB\over 4\pi^2\right) V(\infty)\quad,
\ee
which seems technically much easier. With this formulation, it is not difficult to derive the following results;
\begin{figure}[t]
	\centering
	\includegraphics[width=10cm]{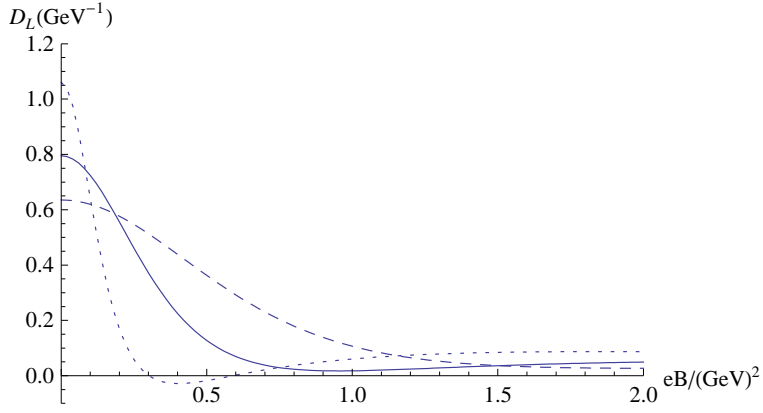}
		\caption{Numerical result for $D_L$ in the Sakai-Sugimoto model with $T=150$ MeV(dotted), $T=200$ MeV(plain), and $T=250$ MeV(dashed). \label{fig2}}
\end{figure}

{\it Weak field limit : $eB\to 0$}
\be
v_\chi \sim \mp\left(N_c eB\over 4\pi^2\right)\int_{U_T}^\infty \,{dU'\over A(U')}+{\cal O}\left(eB\right)^3=\mp{27\over 8\pi^2}{M_{KK}(eB)\over (g_{YM}^2 N_c) T^3}+{\cal O}\left(eB\right)^3\quad.
\ee

{\it Strong field limit : $eB\to\infty$}
\be
v_\chi \to \mp 1\quad({\rm speed\,\, of\,\, light})\quad.
\ee

Note that the strong field limit gives us the same result that one expects in the weak coupling Landau level picture that is discussed in section \ref{section3}. It might come as a surprise because the dynamical degrees of freedom in our holographic model are still mesonic degrees of freedom
represented by world-volume gauge fields on $D8$-branes. They are color singlet, hadronic degrees of freedom, while they reproduce something
related to underlying quark degrees of freedom. This looks quite similar to an idea
of qaurk-hadron duality, and it would be interesting to explore this further.

One can go on to the next order expansion in $k$ to compute the longitudinal diffusion constant $D_L$. As the computation is conceptually
straightforward, we only present our numerical results in our Figure \ref{fig2}. For $eB=0$, it reduces to the known diffusion constant in the model
\be
D_L\to C(U_T)\int_{U_T}^\infty \, {dU'\over A(U')}={1\over 2\pi T} \quad,\quad eB\to 0\quad,
\ee
while in general it is a complicated function of $eB$ and the temperature $T$. 
As one can observe in Figure \ref{fig2}, $D_L$ generally gets decreased as $eB$ increases, which seems physically sensible because a larger magnetic field
would align quasi-particles more efficiently, so that their microscopic longitudinal velocity diffusion would be smaller.
What is interesting is that for some range of $eB$ in a low temperature, say $T=150$ MeV, one seems to have a {\it negative} value of $D_L$.
Looking at the dispersion relation
\be
\omega = \mp v_\chi k - i D_L k^2 +\cdots\quad,
\ee
this signals an instability for sufficiently small $k$ or long wave-length fluctuations, which is precisely similar to Gregory-Laflamme instability in a gravity system \cite{Gregory:1993vy}. In the case of gravity system, Gubser-Mitra conjecture \cite{Gubser:2000ec} links this dynamic instability with a thermodynamic instability, and
it would be interesting to study whether our instability is also related to some kind of thermodynamic instability. We leave this for future studies.

\section{Dynamical Electromagnetism : Mixing Chiral Magnetic Wave with Plasmons} 

The previous sections treat electromagnetism as providing only a non-dynamical external magnetic field, and neglecting dynamical nature
of electromagnetic field. This is valid in the limit $e\to 0$ while keeping $eB$ finite.
However, to describe the real world more precisely, it would be desirable to go beyond this approximation including dynamical electromagnetism.
In general, dynamical electromagnetic field in the plasma couples to longitudinal fluctuations of vector charge density inducing the plasma waves.
The chiral magnetic wave (more precisely its projection onto the vector $U(1)_V$-part) also involves longitudinal charge density fluctuations; 
it is thus natural to expect them to mix with each other, resulting in interesting modifications of their dispersion relations. 
This will be the main topic of this section.
We also point out that for non-Abelian $SU(N_F)$ chiral magnetic waves there is no mixing with plasmons
as far as one does not introduce gauging of $SU(N_F)$, so that the previous discussion stays intact for non-Abelian CMWs.\footnote{To be more precise, EM charge is a sum of $U(1)$ and $I_3$ of $SU(2)_F$ for $N_F=2$, so the situation can be more complicated in general. We leave a full discussion as a simple extension of ours to the readers.}  

Having dynamical electromagnetism, one first needs to include the Maxwell equation as a new dynamical equation of motion 
in addition to the conservation laws of currents,
\bear
\partial_\mu\left(\partial^\mu A^\nu- \partial^\nu A^\mu\right) = e j^\nu_V\quad.\label{maxwell}
\eear
What is important for us is how to determine the current $j_V^\mu$. Typical discussions of plasma waves 
proceed by assuming a linear retarded response of the plasma to the electromagnetic field $A^\mu$,
\bear
j^\mu_V = e \Pi^{\mu\nu}(\omega, k) A_\nu\quad,\quad k_\mu \Pi^{\mu\nu}=0 \quad,\label{retard}
\eear
where we work in the frequency-momentum space. Combined with (\ref{maxwell}) this would result in a self-contained equation for $A_\mu$ from which one can extract the plasmons. 
Let us review this procedure briefly as we are going to extend it by including the chiral magnetic waves and diffusion terms.
Assuming a definite frequency-momentum $e^{-i\omega t + i k x^1}$, we focus on longitudinal polarizations $(A_0,A_1)$, the only ones where we expect the emergence of plasmons.
Because one has
a gauge freedom of shifting $A_\mu$ by $k_\mu$, one can use this to remove $A_1$ and work with $A_0$ only, which simplifies the analysis significantly.
From (\ref{maxwell}) and (\ref{retard}), one then obtains
\bear
k^2 A^0 = e^2 \Pi^{00}(\omega,k) A^0\quad,\quad k\omega A^0 = e^2 \Pi^{10}(\omega,k)A^0\quad.
\eear
These two equations are in fact equivalent as can be seen using the Ward identity $k_\mu\Pi^{\mu 0}=\omega \Pi^{00}-k\Pi^{10}=0$, and one has a non-trivial solution of $A^0$ only if $(\omega,k)$ satisfies the {\it plasmon dispersion relation},
\bear
k^2=e^2 \Pi^{00}(\omega,k)\quad.
\eear
One expects plasma waves in long wave-length regime $k\to 0$ with finite frequency, so one expands $\Pi^{00}$ in powers of $k^2\over \omega^2$ as
\bear
\Pi^{00}\sim \omega_0^2 {k^2\over \omega^2}\left(1+c_{eff}^2 {k^2\over \omega^2}\right)+\cdots\quad,\label{pi00exp}
\eear
upon which the dispersion relation becomes
\bear
\omega^2= e^2 \omega_0^2 +c_{eff}^2 k^2 +{\cal O}(k^4)\quad;
\eear
this looks like a massive excitation of mass squared $\omega_p^2\equiv e^2\omega_0^2$ with an effective speed of light $c_{eff}$.
The parameter $\omega_0$ is typically of the order of temperature $T$, and the plasma frequency is about $\omega_p \sim eT$ which can be small compared to $T$ in weak-coupling limit; the plasmon thus can be an important ingredient in describing the hydrodynamic properties of the system.

We will have to modify this picture in two aspects for our case of a constant background magnetic field
by including chiral magnetic waves and diffusion terms. Because the magnetic field $B$ is constant while the plasma waves involve only longitudinal electric field fluctuations, we can decouple them, and treat $A_\mu$ from now on as an {\it additional} fluctuation of the gauge field on top of the background magnetic field. 
 Looking at the chiral magnetic effect (\ref{semifinal}), one
first notices that the {\it anomalously} induced spatial currents are not
directly related to retarded response to the gauge field fluctuations $A^\mu$; they are induced from the {\it total} charge densities $j^0_{V,A}$
without any regard to how these charge densities appear. We will discuss these charge densities below. Also, we have to modify (\ref{semifinal})
for finite frequency-momentum by introducing chiral magnetic conductivity $\sigma(\omega,k)$ \cite{Kharzeev:2009pj,Yee:2009vw} because it generally depends on $(\omega,k)$,
\be
\left(\begin{array}{c} \vec j_V \\ \vec j_A\end{array}\right) =  \vec B \alpha  \left(\begin{array}{cc}
0 & \sigma(\omega,k) \\ \tilde\sigma(\omega,k) & 0
\end{array}\right) \left(\begin{array}{c}
j^0_V \\ j^0_A
\end{array}\right)\equiv 
 {N_c e \vec B \alpha\over 2\pi^2}  \left(\begin{array}{cc}
0 & \left(\sigma(\omega,k)\over \sigma_0\right)\\ \left(\tilde\sigma(\omega,k)\over \sigma_0\right) & 0
\end{array}\right) \left(\begin{array}{c}
j^0_V \\ j^0_A
\end{array}\right)
\quad.\label{semifinal2}
\ee
where $\sigma_0={N_c e \over 2\pi^2}$ is the zero frequency-momentum limit we have used before. In general, at finite  $(\omega,k)$, we expect
$\sigma\neq \tilde\sigma$.
Therefore the total spatial currents $\vec j_{V,A}$ will represent the sum of the retarded response to $A^\mu$ given by (\ref{retard}) and the anomalously
induced contribution (\ref{semifinal2}),
\bear
j^1_V&=&e\Pi^{10}A^0+{N_c eB \alpha\over 2\pi^2}\left(\sigma(\omega,k)\over \sigma_0\right)j^0_A -ik D_L j^0_V\quad,\nonumber\\
j^1_A&=&e\Pi^{10}_{AV}A^0 +{N_c eB\alpha \over 2\pi^2}\left(\tilde\sigma(\omega,k)\over \sigma_0\right)j^0_V - ik D_L j^0_A\quad,\label{spatialplasmon}
\eear
where we keep our focus on longitudinal components only, and we also included the diffusion terms proportional to $D_L$.
Note that we also include the induced axial current from response to $A^0$ through $\Pi^{10}_{AV}$ because this term indeed exists in the
presence of background magnetic field $B$ (think of triagle diagram of external $B$ and $A^0$ which couples to axial current).
It is important to keep in mind that the charge densities $j^0_{V,A}$ appearing on the right hand side are {\it total} charge densities
that may come from {\it both} response to $A^\mu$ as well as additional fluctuations due to chiral magnetic effects.
Once we write down (\ref{spatialplasmon}), we don't and can't specify charge densities because they are free up to dynamical equations
of Maxwell equation and current conservation laws. To be more precise, we have three dynamic equations; $\nu=0$-component of Maxwell equation and two current conservation laws for $j^\mu_{V,A}$ (the $\nu=1$ Maxwell equation becomes equivalent to $\nu=0$ once $j^\mu_V$-conservation is imposed due to gauge invariance).
They are homogeneous linear equations in terms of three variables $(A_0,j^0_V,j^0_A)$, so that non-zero solutions exist if and only if the $3\times3$
coefficient matrix has zero determinant. This constraint on $(\omega,k)$ will give us the dispersion relation.

Although it is not necessary, it is convenient to decompose $j^0_V$ as
\bear
j^0_V= e\Pi^{00} A^0 +\delta j^0_V\quad,
\eear
to visualize additional fluctuation $\delta j^0_V$ to the retarded response explicitly.
Let us then write down the three independent dynamical equations mentioned above.
The $\nu=0$-component Maxwell equation is
\be
k^2 A^0 = e^2 \Pi^{00} A^0 + e \delta j^0_V\quad,\label{fineq1}
\ee
while the vector current conservation, $\partial_\mu j^\mu_V=0$, looks as
\be
-i\omega\delta j^0_V +ik{N_c eB\alpha \over 2\pi^2}\left(\sigma\over \sigma_0\right)j^0_A +k^2 D_L \left(e\Pi^{00} A^0 + \delta j^0_V \right)=0\quad.
\label{fineq2}
\ee
Finally, for axial current conservation, there is an important modification to its conservation due to
triangle anomaly we are considering (note that we are still neglecting QCD anomaly from gluons). 
Recall that axial current becomes anomalous in the presence of non-zero electromagnetic
$\vec E\cdot \vec B \neq 0$ due to triangle anomaly,
\be
\partial_\mu j^\mu_A = {e^2 N_c\over 16 \pi^2} \epsilon^{\mu\nu\alpha\beta} F_{\mu\nu} F_{\alpha\beta}={e^2 N_c \over 2\pi^2} \vec E\cdot \vec B\quad.
\label{axialano}
\ee
Remember that our $N_c$ quarks have charge $e$ in this paper.
We already have a background magnetic field $\vec B= B \hat x^1$, while dynamical longitudinal plasma fluctuations we are considering
have a longitudinal electric field fluctuation $E_1=\partial_1 A^0=ik A^0$ in Fourier space, so that one has locally non-vanishing 
$\vec E\cdot\vec B$ that affects axial current conservation law as in (\ref{axialano}). The resulting (modified) conservation law of $j^\mu_A$ gives us
\be
-i\omega j^0_A +ik e\Pi^{01}_{AV}A^0+ik{N_c eB\alpha\over 2\pi^2}\left(\tilde\sigma\over \sigma_0\right)\left(e \Pi^{00} A^0 +\delta j^0_V\right) +k^2 D_L j^0_A = ik{N_c e^2 B\over 2\pi^2} A^0\quad,
\label{fineq3}
\ee
where the right hand side is the anomalous contribution that we discussed.
The above equations (\ref{fineq1}),(\ref{fineq2}), and (\ref{fineq3}) are the main equations for $(A^0,\delta j^0_V,j^0_A)$ from which one can obtain
dispersion relations.

As an easy application as well as an illustration, let us turn off the magnetic field for a moment and consider the diffusion effects only, which may be called
{\it diffusive plasmons}.
In this case, one has $\Pi^{10}_{AV}=0$ due to $B=0$, and axial current decouples with the usual diffusion $\omega = -i D_L k^2$, while $(A^0,\delta j^0_V)$ system becomes
\bear
\left(k^2-e^2 \Pi^{00}\right)A^0 &=& e \delta j^0_V\quad,\nonumber\\
\left(i\omega -k^2 D_L\right)\delta j^0_V &=& k^2 e D_L \Pi^{00} A^0\quad,
\eear
which has non-zero solutions if and only if
\be
\left(k^2-e^2 \Pi^{00}(\omega,k)\right)\left(i\omega-k^2 D_L\right)= k^2 e^2 D_L \Pi^{00}(\omega,k)\quad,
\ee
which gives the dispersion relation. Upon expanding $\Pi^{00}$ as in (\ref{pi00exp}), one can solve the above for small $k$ as
\be
\omega^2 = \omega_p^2 +\left(c_{eff}^2 - i \omega_p D_L\right) k^2 +{\cal O}(k^4)\quad,
\ee
where $\omega_p=e\omega_0$ is the plasma frequency.

Going back to our interesting case of non-zero magnetic field $B\neq 0$,
it is straightforward to study (\ref{fineq1}),(\ref{fineq2}), and (\ref{fineq3}) in complete generality, but we will restrict ourselves to the 
case with $D_L=0$ for simplicity in this paper, leaving their full analysis including $D_L$ to the future.
The system then becomes
\bear
k^2 A^0 &=& e^2 \Pi^{00} A^0+e \delta j^0_V\quad,\nonumber\\
 -i\omega\delta j^0_V +ik {N_c eB\alpha\over 2\pi^2} \left(\sigma \over \sigma_0\right)j^0_A &=& 0 \quad,\nonumber\\
-i\omega j^0_A +ike\Pi^{10}_{AV}A^0+ ik{N_c eB\alpha\over 2\pi^2}\left(\tilde\sigma\over \sigma_0\right) \left(e\Pi^{00} A^0+\delta j^0_V\right) &=& ik{N_c e^2 B\over 2\pi^2} A^0\quad,
\eear
which mixes all three fluctuations together. From the first equation, one can replace $A^0$ with $\delta j^0_V$, and inserting it into the other
two equations, one gets 
\bear
-i\omega\delta j^0_V +ik{N_c eB \alpha\over 2\pi^2} \left(\sigma\over \sigma_0\right)j^0_A &=&0\quad,\nonumber\\
-i\omega j^0_A+ik {N_c eB \alpha\over 2\pi^2}\left(\tilde\sigma\over \sigma_0\right){\left(k^2-{e^2\over \alpha}{\sigma_0\over \tilde\sigma}+
{2\pi^2 e\over N_c B\alpha }{\sigma_0\over\tilde\sigma}\Pi^{10}_{AV}\right) \over k^2-e^2 \Pi^{00}} \delta j^0_V &=& 0\quad,
\eear
from which one gets the dispersion equation 
\be
\omega^2 = v_\chi^2 k^2 \left(\sigma(\omega,k)\over\sigma_0\right)
\left(\tilde\sigma(\omega,k)\over\sigma_0\right){\left(k^2-{e^2\over \alpha}{\sigma_0\over\tilde\sigma(\omega,k)}+{2\pi^2 e\over N_c B\alpha}{\sigma_0\over\tilde\sigma(\omega,k)}\Pi^{10}_{AV}(\omega,k)\right) \over k^2-e^2 \Pi^{00}(\omega,k)}\quad,\label{plasmoncmw}
\ee
where $v_\chi={N_c eB\alpha\over 2\pi^2}$ as before. This equation is our master equation that governs mixing between chiral magnetic waves
and plasma waves.

For small magnetic field $B$ and $\omega\sim \omega_p\sim e T \ll T$, one expects that chiral magnetic conductivities are approximately the zero-frequency value $\sigma\approx\tilde \sigma\approx\sigma_0$, and moreover anomaly triangle diagram gives us
\bear
\Pi^{10}_{AV}(\omega,k) \to {N_c eB\over 2\pi^2} \quad {\rm as} \quad \omega,k\to 0\quad,
\eear
so that the numerator in the right hand side of (\ref{plasmoncmw}) becomes simplified.
One  then uses the previous expansion (\ref{pi00exp}) of $\Pi^{00}$ to solve the above equation to get
\bear
\omega^2= \omega_p^2 +\left(v_\chi^2+c_{eff}^2\right) k^2 +{\cal O}(k^4)\quad,
\eear
which describes effects from chiral magnetic wave to the plasma waves. Note that the effect exists only with finite $k$, and this makes sense
because chiral magnetic waves disappear in $k\to 0$ limit.

Another interesting limit is an infinitely large $B\to\infty$ limit, where one expects effective reduction to 1+1 dimensional theory.
In fact, making electromagnetism dynamical corresponds to 1+1 dimensional QED with $N_c$ massless Dirac fermions, or the Schwinger model \cite{Schwinger:1962tp}.
It has been known for long time that the photon in the model becomes massive due to 1+1 dimensional axial anomaly,
\be
m^2_\gamma = {N_c e^2_{eff}\over\pi}\quad,
\ee
where $e^2_{eff}$ is an effective 1+1 dimensional QED coupling constant.
As the 4-dimensional triangle anomaly (\ref{axialano}) correctly 
reduces to 1+1 dimensional axial anomaly in the presence of background magnetic field $B$, 
one should be able to reproduce this Schwinger phenomenon from our master equation (\ref{plasmoncmw}) in the limit $B\to\infty$.

One can be more quantitative to test this connection.
To find $e_{eff}^2$, it is useful
to consider a transverse area of 
\be
\int d^2 x_T = {2\pi\over eB}\quad,
\ee
to have a single lowest Landau level system per each 4D fermion because the transverse density of LLL is $eB/2\pi$.
Thinking of fermion kinetic term, the proper normalization between 4D fermion and 2D fermion is
\bear
\psi_{4D}=\sqrt{eB\over 2\pi} \psi_{2D}\quad,\label{nor}
\eear
which will be useful shortly when we discuss about $\Pi^{00}$.
Because one is looking at only longitudinal dynamics of 4D $U(1)$ gauge field, the only relevant dynamical field is $F_{01}$, and
the gauge field action indeed reduces to 1+1 dimensional QED action as
\bear
{1\over 2 e^2}\int d^4 x \, \left(F_{01}\right)^2 ={1\over 2 e^2}{2\pi\over eB}\int d^2x\, \left(F_{01}\right)^2\equiv 
{1\over 2 e^2_{eff}}\int d^2x\, \left(F_{01}\right)^2\quad,
\eear
so that $e_{eff}^2= {e^3 B\over 2\pi}$, and the Schwinger photon mass should be
\bear
m_\gamma^2 = {N_c e^3 B\over 2\pi^2}\quad.
\eear
To reproduce this from our equation (\ref{plasmoncmw}), note that as $B\to\infty$, the expected $\omega^2=m_\gamma^2$ is also infinite
and one naturally expects that chiral magnetic conductivities go to zero in this limit as the system cannot respond to arbitrary fast
perturbations; $(\sigma,\tilde\sigma)\to 0$ as $\omega\to \infty$. Therefore, the solution of (\ref{plasmoncmw}) in this limit
is found simply by demanding that the denominator vanishes or
\bear
k^2=e^2 \Pi^{00}(\omega,k)\quad,\label{11}
\eear
where $\Pi^{00}$ should be given by the effective 1+1 dimensional theory. To find it, recall from (\ref{nor}) that $j_{4D}^\mu={eB\over 2\pi}j_{2D}^\mu$, so that
\bear
\Pi^{00} \sim \int d^4 x e^{-i\omega t +ikx}\langle j^0_{4D}(x) j^0_{4D}(0) \rangle 
={2\pi\over eB} \left(eB\over 2\pi\right)^2 \int d^2 x e^{-i\omega t +ikx}\langle j^0_{2D}(x) j^0_{2D}(0) \rangle 
=\left({eB\over 2\pi}\right)\Pi^{00}_{2D}.\nonumber\\
\eear
As we have $N_c$ 1+1 dimensional Dirac fermions, $\Pi^{00}_{2D}$ is $N_c$ times that of a single Dirac fermion, which can be found most easily by bosonization to
a single real scalar field $\phi$ such that
\be
j^{\mu}_V = \sqrt{1\over\pi}\epsilon^{\mu\nu}\partial_\nu \phi\quad,\quad j^\mu_A=\sqrt{1\over\pi} \partial^\mu \phi\quad,
\ee
where $\phi$ is normalized to have a standard kinetic term ${\cal L}={1\over 2}\partial_\mu\phi \partial^\mu \phi$.
Then, $\Pi^{00}_{2D}$ is easily computed as
\be
\Pi^{00}_{2D} ={N_c\over\pi} \langle (\partial_1\phi) (\partial_1\phi)\rangle= {N_c\over \pi}{(ik)(-ik) \over \omega^2-k^2 }={N_c \over\pi}{k^2\over\omega^2-k^2}\quad,
\ee
so that the equation (\ref{11}) becomes
\be
k^2= e^2 \Pi^{00}_{4D}= {N_c e^3 B\over 2\pi^2}{k^2\over \omega^2-k^2}\quad,
\ee
which indeed gives us $\omega^2=m_\gamma^2+k^2$ reproducing the Schwinger model result.

Therefore the plasmon in the dimensionally reduced theory can be seen as a result of the interaction of the dynamical photon with the chiral magnetic waves.

\section{Summary \label{section5}}

We have demonstrated that the Chiral Magnetic and Chiral Separation Effects (CME and CSE) in relativistic plasmas subjected to magnetic field imply the existence of a new type of a collective excitation in the plasma - the Chiral Magnetic Wave (CMW). This excitation represents the density waves of electric and chiral charge coupled by the triangle anomaly. In strong magnetic field the CMW propagates with the velocity of light, $v_\chi \to 1$. In weak magnetic field, the velocity $v_\chi$ decreases; the result of the holographic computation is shown in Figure \ref{fig1}.  At weak coupling, this decrease of the velocity of the CMW can be understood as originating from the admixture of the excited Landau levels.

The existence of CMW in the quark-gluon plasma has important implications for the phenomenology of heavy ion collisions. The CME relies on the fluctuation of the axial charge density and so the net effect is expected to vanish when averaged over many events; one thus relies on measuring the fluctuations of charge asymmetries \cite{Kharzeev:2004ey,Voloshin:2004vk}. On the other hand, the CMW should exist even in a neutral plasma, and so does not require the presence of the axial or baryon chemical potentials. 
Since it represents the coupled density waves of electric and chiral charges propagating along the direction of the applied magnetic field (that in heavy ion collisions is perpendicular to the reaction plane), the CMW can induce dynamical, reaction plane dependent, fluctuations of electric charge. The azimuthal angle dependence of these fluctuations will be determined by the wavelength of the CMW excitation. We will return to the consideration of phenomenology related to CMW in heavy ion collisions in a forthcoming publication \cite{jinfeng}.

\vskip 1cm \centerline{\large \bf Acknowledgement} \vskip 0.5cm
We thank G. Ba\c sar, Y. Burnier, G. Dunne, C. Herzog, J. Liao, R. Pisarski, E. Shuryak, D. Son, D. Teaney and I. Zahed for useful discussions. The work of D.K. was supported in part by the U.S. Department of Energy 
under Contract No.~DE-AC02-98CH10886. The work of H.U.Y. was supported by the U.S. Department of Energy under Contract No.~DE-FG02-88ER40388.

\vskip 1cm {\bf Note added:}
After this paper was submitted, we learned of an interesting paper by G. Newman \cite{Newman:2005hd}. In the section VI of that paper, the author arrives at the results similar to ours in section II. We thank D. Son for bringing this work  to our attention.

 \vfil

\end{document}